\documentclass[prl,twocolumn,letterpaper,superscriptaddress]{revtex4}

\usepackage[utf8x]{inputenc}
\usepackage[pdftex]{graphicx}
\usepackage{amsmath,amssymb}
\usepackage{subfigure}
\usepackage{todonotes}
\usepackage{bbold}

\newlength{\graphwid} 

\begin{document}

\title{Nonperturbative and non-Markovian 
F\"{o}rster-interaction in waveguiding systems}

\author{T. Sproll} 
\affiliation{Max-Born-Institut, 12489 Berlin, Germany}

\author{Ch. Martens} 
\affiliation{Max-Born-Institut, 12489 Berlin, Germany}

\author{M. P. Schneider} 
\affiliation{Max-Born-Institut, 12489 Berlin, Germany}

\author{F. Intravaia} 
\affiliation{Max-Born-Institut, 12489 Berlin, Germany}
\affiliation{Humboldt-Universit\"at zu Berlin, Institut f\"ur Physik,  AG Theoretische Optik \& Photonik, 12489 Berlin, Germany}

\author{K. Busch} 
\affiliation{Max-Born-Institut, 12489 Berlin, Germany} 
\affiliation{Humboldt-Universit\"at zu Berlin, Institut f\"ur Physik,  AG Theoretische Optik \& Photonik, 12489 Berlin, Germany}


\begin{abstract}
\noindent 
An exact solution to the interaction between two emitters mediated by F\"{o}rster resonance energy 
transfer is presented. The system is comprised of a one-dimensional optical waveguide with two embedded 
two-level systems and is analyzed using a recently developed quantum-field theoretical approach. This 
exactly solvable model features several competing mechanisms that lead to a rich physical behavior such
as, for instance, the possibility to change from an attractive to a repulsive interaction. 
Our nonperturbative analysis allows on very general grounds to describe the interaction as a truly 
non-Markovian phenomenon mediated by the atom-photon bound states of the system. 
These results are of direct relevance for energy or information transfer processes as well as for atom 
trapping more complex systems. 
\end{abstract}

\maketitle

Since the seminal paper by T. F\"{o}rster in 1948 \cite{Forster48}, the field of nonradiative resonant 
energy transfer (FRET) has found wide-ranging applications in physics, chemistry and biology \cite{Sekar03,Saini06}. 
In FRET, energy is exchanged between two emitters through nonradiative dipole-dipole interaction.
This phenomenon is used to obtain information about signal transport inside biological systems 
\cite{Endo10} as well as to study the behavior and the interaction between proteins \cite{Pollock99,Truong01}. 
Several methods of optical microscopy have used dimer-based FRET-induced fluorescence to circumvent 
the Abb\'{e} diffraction limit \cite{Ishikawa12}. 
On another front, over the past decade, waveguide quantum electrodynamics (WQED) has attracted a growing
attention \cite{Longo11,Zheng13,Laakso14,Shahmoon14,Hood16,Shi16,Lodahl17}. In WQED systems, quantum emitters 
are coupled to an essentially one-dimensional electromagnetic environment, leading to interesting 
applications in the context of quantum information processing~\cite{Ladd10}, also including the 
transport and storage of quantum information (often termed as the ``Quantum Internet''~\cite{Kimble08}). 
Physical realizations at optical frequencies range from fibers with nearby trapped cold atoms~\cite{Reitz13} 
to photonic crystal waveguides with embedded emitters~\cite{Arcari14}. 

Here, we bring together these subjects of growing significance in physics and study their interplay. 
We show that for FRET in WQED systems, dispersion relation effects (e.g. slow light \cite{Longo11}) 
are of paramount importance. Specifically, at the level of the light-matter interaction, the peculiar 
behavior of the electromagnetic group-velocity, which under certain circumstances can be reduced to 
almost zero, introduces significant memory effects, carrying an intrinsic signature of non-Markovianity, 
which strongly modify the dynamic of the system. 
Probably the most unexpected manifestation of these phenomena is the appearance in the emitter's dressing 
process of states, often termed atom-photon bound states (APBSs) \cite{John90}. Since these states 
drastically affect the nonradiative properties of the atom, they have a strong impact on the physics
of the FRET mechanism.
Our interest is twofold. First and contrary to many earlier investigations, the simplicity of our model 
system allows to discuss FRET in a controlled environment, leading to a deeper understanding and to the
highlighting of the relevant physical mechanisms. 
Second, FRET and related phenomena provide novel interesting effects and opportunities for WQED, such 
as additional communication channels and distinctive physical interactions mechanisms (e.g. trapping 
potentials).
\begin{figure}
\centering
	  \includegraphics[width=0.45\textwidth]{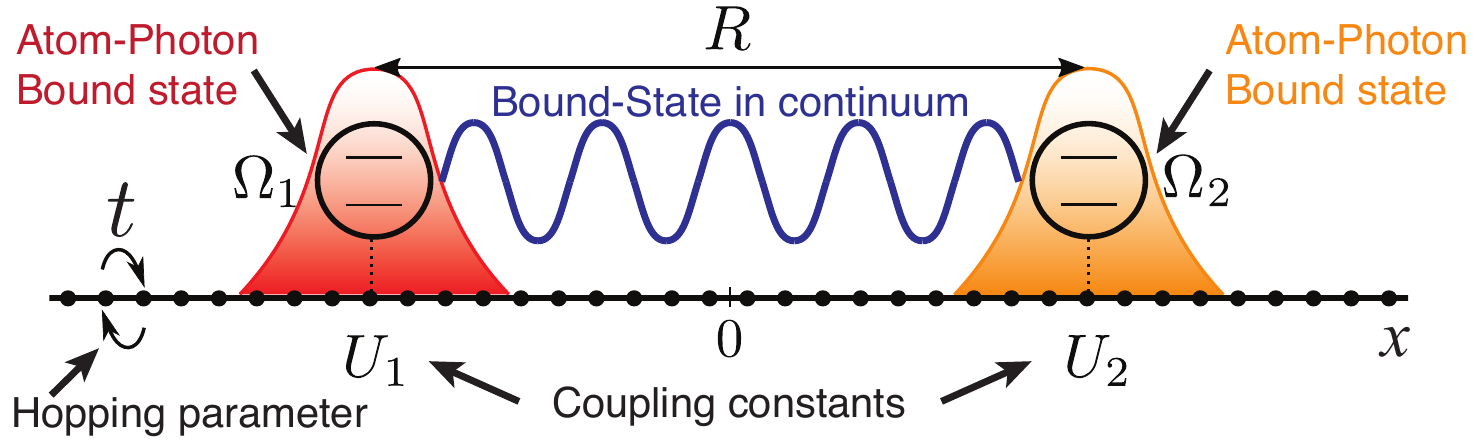}
	\caption{(Color online) Schematic of the model system for studying FRET within WQED. Two TLSs with 
	         frequencies $\Omega_{1,2}$ coupled with strength $U_{1,2}$ to a waveguide with dispersion 
					 relation $\epsilon(k)$ and hopping parameter $t$.\label{22levelsystems}}
\end{figure}

\begin{figure}
	 \includegraphics[width=0.45\textwidth]{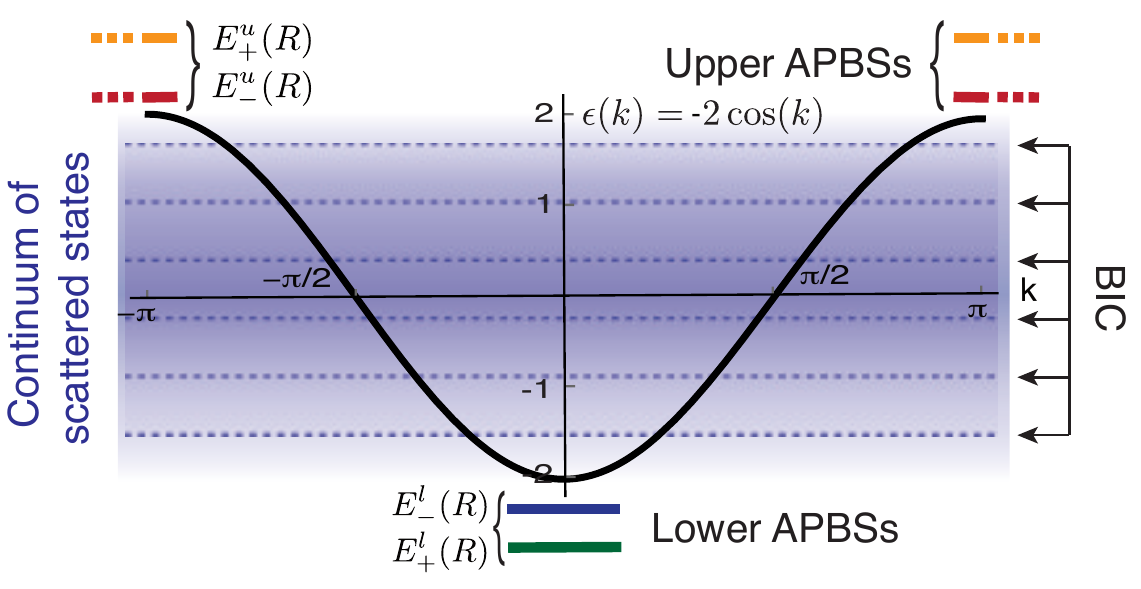}
	 \caption{(Color online) Energetic configuration for the Hamiltonian in Eq. \eqref{eq:Hamiltonian} 
	          with  $\epsilon(k)=-2 \cos(k)$ (the energies are normalized to the hopping parameter 
						$t$; see Fig. \ref{22levelsystems}).
	          The system's eigenenergies fall into three classes: a continuum of scattering states (shadowed area), bound states
						in the continuum (BICs) \cite{Note1} and atom-photon bound states (APBSs). 
	 \label{EnergySpectrum}}
\end{figure}

Our system consists of a one-dimensional bosonic quantum wire formed by a chain of sites coupled to two 
two-level-systems (TLSs) separated by a distance $R$ (measured in units of site spacing, see Fig.~\ref{22levelsystems}). 
To make the analysis of the underlying physical processes more transparent it is useful to describe 
each TLS within a slave fermion representation \cite{Auerbach94,Schneider16}. Then, the Hamiltonian 
of our model system is given by ($\hbar=1$)  
\begin{align}
   \hat{H} & =    \sum_k \epsilon(k) \hat{a}^{\dagger}_k \hat{a}_k 
	              + \sum_{j=1}^2 \frac{\Omega^{\{j\}}}{2} 
								               \left[\hat{f}^{\dagger \{j\}} \hat{f}^{\{j\}} - \hat{g}^{\dagger \{j\}} \hat{g}^{\{j\}}\right] 
				        \nonumber \\
           & \phantom{=}
		            + \sum_k\sum_{j=1}^2 \left[U_j e^{i s(j) k R/2} \hat{a}^{\dagger}_k \hat{f}^{\{j\}} \hat{g}^{\dagger \{j\}} + h.c. \right]
		\label{eq:Hamiltonian}	
\end{align}
Here, $\hat{a}^{(\dagger)}_k$ denotes photonic annihilation (creation) operators for the modes 
with wave number $k$ of a waveguide with dispersion relation $\epsilon(k)$. Likewise, $\hat{f}_j^{(\dagger)}$ 
and $\hat{g}_j^{(\dagger)}$ represent fermionic annihilation (creation) operators corresponding to the 
ground ($\hat{g}_j$) and excited states ($\hat{f}_j$) of the two TLSs (labeled by $j=1,2$) with 
corresponding level separations $\Omega_j$. 
The constant $U_j$, describing the coupling strength of the TLSs to the waveguide can, without
loss of generality, be assumed to be real and $s(j)=(-1)^j $ determines the sign in the Peierls 
factors \cite{Schneider16}. In addition, we have employed the rotating wave approximation and the 
dipole approximation which is well-justified in the optical regime \cite{Mercouris96,Feranchuk96}.
For FRET, we can restrict ourselves to the one-excitation sector of the combined atom-photon 
Hilbert space. Here the (dressed) eigenstates of the Hamiltonian in Eq. \eqref{eq:Hamiltonian} 
naturally separate into three classes (see Fig. \ref{EnergySpectrum}). The first class corresponds
to scattering states, which exhibit frequencies within the spectral bandwidth of the original 
waveguide.
Bound states in the continuum (BIC)~\cite{Stillinger75}, characterized by discrete energies 
located within a dense continuum of scattering states form the second class.
These BICs show up for identical atoms and correspond to the geometry-induced states of a 
perfect cavity where the two TLSs act as perfect mirrors~\cite{Note1,Zumofen08,Longo11,Sproll16}.
Finally, we have the the third class of APBSs, whose eigenenergies are situated outside the spectrum 
of the waveguide \cite{John90} (see Fig. \ref{EnergySpectrum}). They are characterized by wave 
functions that are exponentially localized around the locations of the TLSs \cite{Longo11}. 
APBSs only occur when the waveguide dispersion relation exhibits frequency cut-offs or band edges 
that are associated with slow-light regimes and non-Markovian dynamics \cite{John90,Economou05,Liu17}. 
The frequencies of the scattering states are insensitive to the separation of the TLSs and as 
the BIC states are embedded in a continuum, they induce only a small interaction between the
TLSs which we shall neglect in subsequent discussion. 
Conversely, the number and properties of the APBSs strongly depend on the TLS separation $R$, 
inducing a significant nonradiative interaction between the emitters. 

Using the properties of these states we can describe the waveguide-mediated interaction between 
the two TLSs through the F\"{o}rster potential defined as follows
\begin{equation}
 	\phi_{|\Psi\rangle}(R) = \sum_{i} \left[ \alpha_{|\Psi\rangle,i}(R) E_i(R)-\alpha_{|\Psi\rangle,i}(\infty) E_i(\infty) \right].
	 \label{FoersterDef}
\end{equation}
In Eq. \eqref{FoersterDef}, $E_i(R)$ denote the eigenenergies of the APBSs, $|\Psi\rangle$ is 
the state in which the system is initially prepared and $\alpha_{|\Psi\rangle,i}=|\langle \Psi| E_i \rangle|^2$ 
represent the eigenstate's occupation numbers. Within our model, the coefficients $\alpha_{|\Psi\rangle,i}$ 
play the same role as the orientation factors of ordinary three-dimensional FRET theory \cite{Forster48}. 
Notice that, due to the properties of the APBSs, the above expression can be regarded as a direct 
consequence of the non-Markovian dynamics of our system and looses its meaning as soon as these 
features disappear (e.g. for a linear waveguide dispersion relation, see below). Equation \eqref{FoersterDef}
reminds the definition of the Casimir energy \cite{Casimir48} or, more specifically, its 
polaritonic contribution \cite{Intravaia05,Intravaia07}. Physically, it describes how the system's 
contribution to energy that stems from the APBSs, 
changes as a function of the distance between the TLSs. As for the Casimir effect, the $R\to \infty$ 
limit appearing in Eq. \eqref{FoersterDef} sets the zero of the interaction potential to the 
configuration where the TLSs are well separated \cite{Intravaia12b}.
Note, however, that since in our case at least one of the TLSs is excited, 
we are dealing with an interaction that is similar to the van der Waals-Casimir-Polder potential of 
an excited atom near a surface \cite{Wylie85,Intravaia11,Laliotis14} or near another atom \cite{Haakh12b,Donaire15a,Milonni15}.

\begin{figure*}
  	\includegraphics[width=0.95\textwidth]{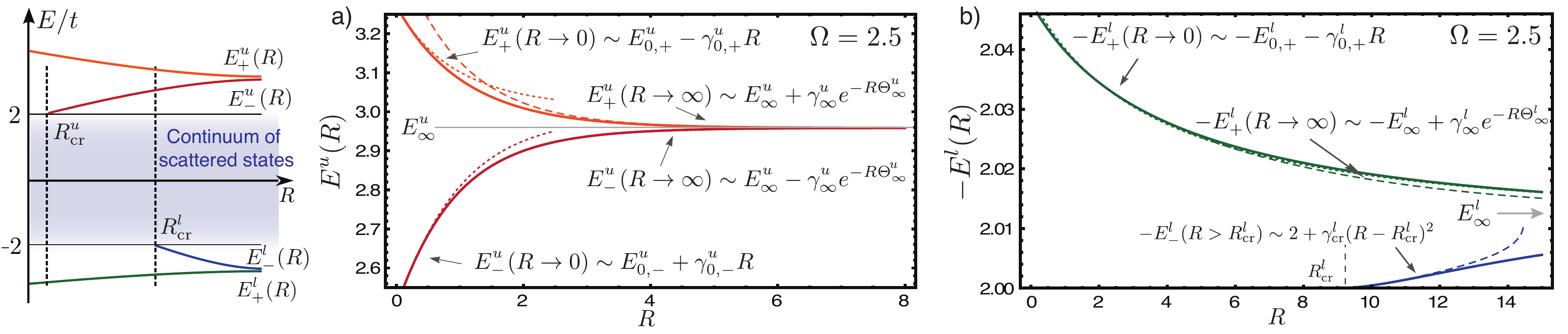}
	\caption{(Color online) Eigenergies of the APBSs for a system with cosine-shaped dispersion relation. 
	         Left panel: A sketch of the typical behavior of the solutions of Eq. \eqref{eigenvaluecos}. 
					 Panels a) and b): Numerical solutions of Eq. \eqref{eigenvaluecos} (full line) and the 
					 corresponding asymptotic calculations \cite{SuppMat}. 
					 In both graphs, $E^{u,l}_{0,\pm}=2\cosh(\Theta^{u,l}_{0,\pm})$ and $E^{u,l}_{\infty} = 2\cosh(\Theta^{u,l}_{\infty})$ 
					 denote the eigenvalues of Eq. \eqref{eigenvaluecos} at $R=0$ and $R=\infty$, respectively. 
					 We have introduced the quantities $\gamma^{u,l}_{0,\pm} = D( \Theta^{u,l}_{0,\pm})$  and  
					 $\gamma^{u,l}_{\infty}= D( \Theta^{u,l}_{\infty})$, where $D( \Theta) =U^2 \sinh(\Theta)/(2[\cosh(2 \Theta)-\Omega\cosh(\Theta)])$, 
					 and $\gamma^{l}_{\rm cr}=4/(R^{l}_{\rm cr})^2$ \cite{SuppMat}.		
           Panel a): Energy values above the band ($E^{u}_{+,0}\approx 3.243$ $E^{u}_{-,0}\approx 2.544$, 
					 $E^{u}_{\infty}\approx 2.959$) with parameters $\Omega=2.5$ and  $U=1$. Note that both 
					 curves stay finite for $R = 0$. Panel b): Negative of the energy values below the band 
					 ($E^{l}_{+,0}\approx -2.046$, $E^{l}_{\infty}\approx-2.012$) with the same parameters. 
					 The black line in panel b) indicates the critical value $R_{\rm cr}$. See text for further 
					 details.}
	\label{EigVal}
\end{figure*}

In order to calculate $E_i(R)$ from the Hamiltonian \eqref{eq:Hamiltonian}, we employ and extend a 
Feynman diagram technique which has recently been developed and applied to the case of a single 
TLS~\cite{Schneider16} (see also~\cite{Kocabas16}). The spectrum of a Hamiltonian can be obtained 
via the poles of the system's Green function~\cite{Mahan00}.
In our subsequent calculations we focus on a cosine-shaped dispersion $\epsilon(k)=-2 \cos(k)$ which 
corresponds to a one-dimensional tight-binding chain with unit lattice spacing \cite{Economou05} 
(hereafter, all energies and frequencies are normalized to the hopping parameter, $t$ in Fig.~\ref{22levelsystems}). 
This dispersion relation describes generic features such as band edges and slow-light regimes~\cite{Longo11}. 
Both of them impact on the dynamics and the energy spectrum of the system, as it can be seen by 
inspecting first the self-energy $\Sigma_{j}(E) = -i \pi U^2_j \rho(E)$ of an isolated TLS within 
a waveguide \cite{Schneider16}.
For a linear dispersion relation $\epsilon(k)\propto\chi v k$, with chirality $\chi$ and propagation 
speed $v$, the self-energy is constant since the density of states is $\rho(E) = 1/(\pi v)$. In 
this case, the Weisskopf-Wigner approximation~\cite{Weisskopf30} (equivalent to the Markov 
approximation~\cite{Berman10}) holds exactly and the system undergoes a completely Markovian time 
evolution. In contrast, for the cosine-shaped dispersion relation the occurrence of a square-root 
singularity in the density of states $\rho(E)=i/[\pi\sqrt{E^2-4}]$ indicates a strongly non-Markovian 
behavior, especially for dynamics with energies located at borders of the waveguide energy-band, i.e. 
when $|E|\sim 2$. 
Physically, this can be understood as resulting from the long atom-photon interaction times occurring 
in the slow light regimes and characterizing the system (memory-sensitive) time evolution \cite{Longo11}.

An analogous but considerably richer behavior occurs in a system that contain two TLSs, which for 
simplicity, we take to be identical in the remainder of this work ($U_j=U, \Omega_j=\Omega$). 
Upon setting up the Feynman rules and summing up the resulting Dyson equation~\cite{Schneider16,Sproll16,SuppMat}, 
we determine the Green's tensor of the system
\begin{equation}
	\label{GeneralGreensfunction}
	\underline{\mathcal{G}}(R,E,k) = \underline{\mathcal{M}}(R,E,k)/Q(R,E).
\end{equation}
In this expression $Q(R,E)$ is a scalar function (see also Eq.~\eqref{GeneralResCond} below) and 
$\underline{\mathcal{M}}(R,E,k)$ is a second rank tensor which component-wise is analytic 
in $R$~\cite{SuppMat}. This means that all the nontrivial eigenenergies are determined by $Q(R,E)=0$, 
which leads to the equation
\begin{equation}
	\label{GeneralResCond}
	(E-\Omega- \Sigma(E))^2- \Gamma^2(R,E) = 0. 
\end{equation}
In this equation, $\Gamma(R,E)=U^{2}\int G^0_w(k)e^{(-1)^{s(j)} i k R}\mathrm{d}k$ represents a 
generalized self-energy ($\lim_{R\rightarrow0}\Gamma(R,E)=\Sigma(E)$) and $G^0_w=[E-\epsilon(k)+i0^{+}]^{-1}$ 
is the free-waveguide Green function. 
The eigenenergies $E_i(R)$ are real-valued solutions of Eq.~\eqref{GeneralResCond} 
\cite{Note1}.
Since the APBSs' eigenenergies lie outside the photon continuum ($|E| >2$), it is convenient to rewrite Eq. \eqref{GeneralResCond} as
\begin{align}
	\label{eigenvaluecos}
  2 \sinh(\Theta) \left( 2 \cosh(\Theta)-\Omega \right) \, \,\, = U^2 \left( 1\pm e^{-\Theta R} \right),
\end{align}
where we have introduced the parametrization $\Theta(E)=\text{arccosh}(E/2) +\mathrm{he}(-E) i \pi$, 
where $\mathrm{he}(z)$ denotes the Heaviside function. In general, Eq.~\eqref{eigenvaluecos} features 
four distinct solutions: The $\pm$ distinguishes the two solutions found above ($E^{u}_{\pm}>2$) and 
the two solutions found below ($E^{l}_{\pm}<-2$) the cosine band (see Fig. \ref{EigVal} for their 
behavior and corresponding asymptotic expressions \cite{SuppMat}).
However, some of these solutions exist only if the condition
\begin{equation}
	\label{NumberCond}
	R > R^{u,l}_{\rm cr} = 2(2-\text{sign}(E^{u,l})|\Omega|)/U^{2}>0
\end{equation}
is fulfilled. Equation \eqref{NumberCond} indicates, that if $R$ is larger than some critical value(s), 
one or, if $|\Omega|<2$, even two (one below, for $R<R_{\rm cr}^{l}$, and one above the band, for 
$R<R_{\rm cr}^{u}$) of the four APBSs disappear into the waveguide continuum (see
Fig. \ref{EigVal})  \cite{Haakh13}. This behavior can be viewed as consequence of the ``level repulsion'' 
occurring in the interacting system and giving rise to binding and anti-binding energetic configurations 
\cite{Intravaia05,Haakh13}. The level repulsion increases for shorter distances and, depending on the 
value of $E$, one or even two of the four levels approach the band edge, eventually merging with the 
continuum of the scattering states at $R=R^{u,l}_{\rm cr}$. 
Fig.~\ref{EigVal} depicts the numerical solutions of Eq. \eqref{eigenvaluecos} for $\Omega=2.5$ and 
$U=1$ as well as the corresponding asymptotic behaviors in the limits $R\rightarrow 0$,$\,R\approx R^{l}_{\rm cr}$ 
(the only existing in this case) and $R\rightarrow \infty$.  
In agreement with Eq. \eqref{NumberCond}, our analysis shows that one of the solutions below the band 
abruptly disappears in the continuum of scattering states and its behavior at $R\to R^{l}_{\rm cr}$ 
is non-analytic in $R$. In fact for our model, as soon as the bound state reaches the band edge the 
associated wave function suddenly spreads over the entire waveguide, leading to a discontinuity in 
the derivative of the energy eigenvalue. 

\begin{figure}[t]
   	\includegraphics[width=0.47\textwidth]{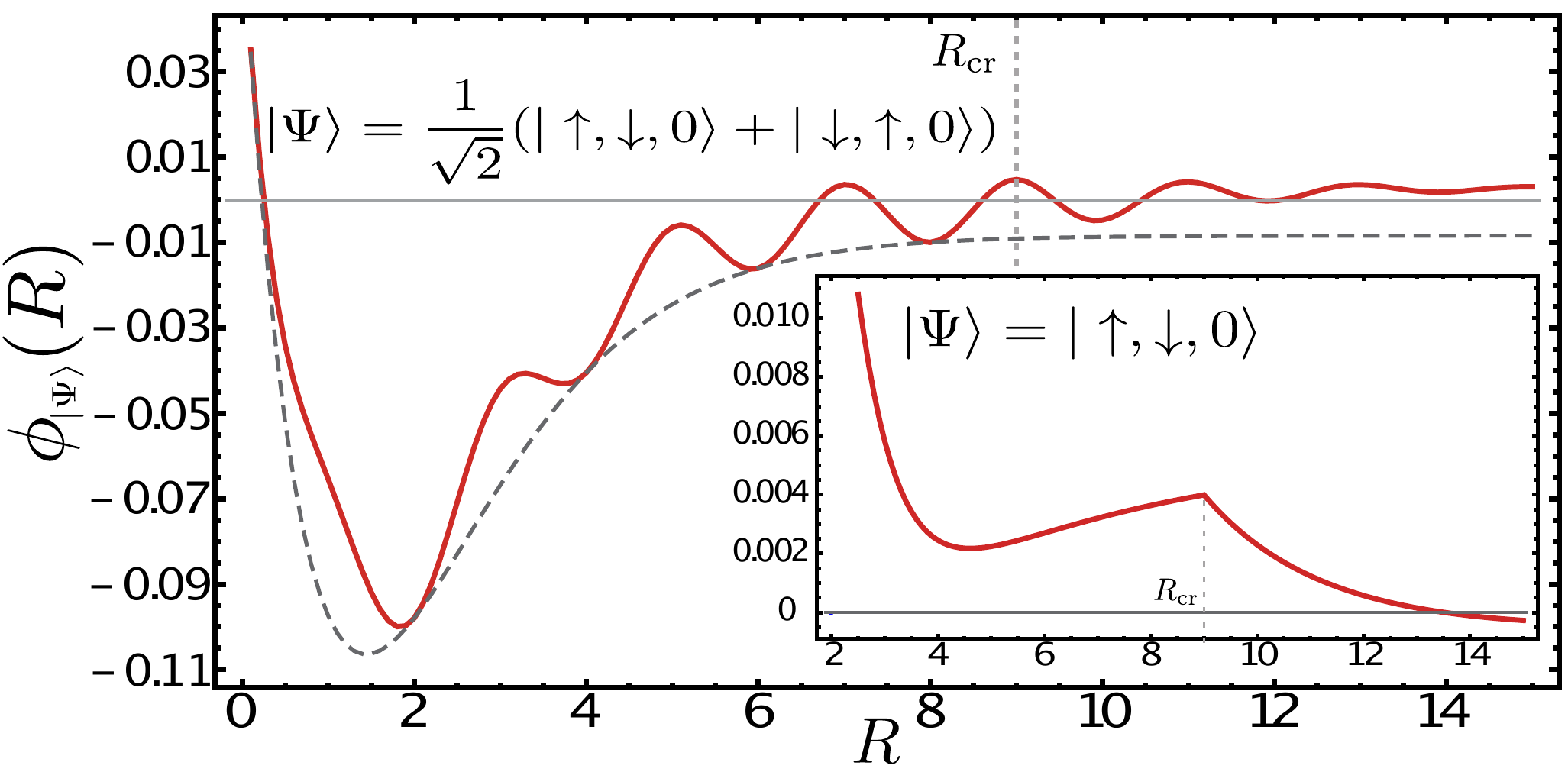}
	\caption{(Color online) F\"{o}rster potential $\phi_{|\Psi\rangle}(R)$ [Eq. \eqref{FoersterDef}] 
	         as a function of the distance between two identical TLS ($\Omega=2.5$) coupled to waveguide (coupling strength $U=1$). 
		       For an initial state $|\Psi\rangle = (|\uparrow,\downarrow,0 \rangle+|\downarrow,\uparrow, 0 \rangle)/\sqrt{2}$, 
					 $\phi_{|\Psi\rangle}(R)$ oscillates due to the change of parity of the APBS when
					 changing $R$. It is dominated by contributions from eigenstates above the photon band 
				 	 (dashed line) while the oscillations come from contributions below the band. 
					 The F\"{o}rster potential exhibits a global minimum at $R \approx 1.5$.
					 Inset: For $|\Psi\rangle=|\uparrow,\downarrow,0 \rangle$, $\phi_{|\Psi\rangle}(R)$ 
					 shows a cusp due to the sudden emergence of a APBS from the continuum of scattering states. }
	\label{FoersterCompPlotTot}
\end{figure}

The last quantities required for calculating the FRET potential in Eq. \eqref{FoersterDef} are the 
occupation numbers $\alpha_{|\Psi\rangle,i}$, which can be evaluated using resolvent theory \cite{Sproll16,SuppMat}. 
For the case of a initial state $|\Psi\rangle \equiv |1\rangle = |\uparrow,\downarrow,0 \rangle$, 
corresponding to one TLS being in the excited state (up arrow), the other TLS being in the ground state 
(down arrow) and zero photons in the waveguide we obtain \cite{SuppMat}
\begin{equation}
	\alpha_{|1\rangle,i}(R) = \text{Res}\{\mathcal{G}_{11}(R,E);E=E_i\},
\end{equation}
where the Green's function $\mathcal{G}_{11}$ accounts for the propagation from the excited TLS back 
to itself including all scattering events.
Upon inserting this information into Eq. \eqref{FoersterDef}, we obtain the F\"{o}rster potential
for two identical TLSs depicted in the inset of Fig. \ref{FoersterCompPlotTot}. 
As we consider transition frequencies $\Omega$ near the upper band edge, the contributions 
from the bound states above the band are much larger in magnitude and decay much faster with
$R$ than their counterparts from below the band. The contributions from below 
the band are negative and lead to a minimum in the potential $\phi_{|1\rangle}(R)$ which, 
for our choice of parameters, can be observed for $R\approx 4.5$.
For $R=R^{l}_{\rm cr}$, the emergence of one of the bound states from the continuum of the 
band leads to the sudden opening of a new nonradiative 
channel which creates a cusp in the F\"{o}rster potential. For even larger values
of $R$, $\phi_{|1\rangle}(R)$ becomes negative: This is due to the contribution 
of the added nonradiative channel which exceeds the contribution of others. 
As the F\"{o}rster potential approaches zero for $R \to \infty$, this results in a shallow 
minimum (not shown in Fig. \ref{FoersterCompPlotTot}).
To investigate the sensitivity of the potential to the initial state of the quantum system, 
it is interesting to calculate the F\"{o}rster potential for the initial state 
$|\Psi\rangle \equiv|\pm\rangle= (|\uparrow,\downarrow,0 \rangle\pm|\downarrow,\uparrow, 0 \rangle)/\sqrt{2}$
(we keep all other parameters the same). Proceeding as before, the corresponding occupation 
numbers above the band are
\begin{equation}
  \alpha^{u}_{|\pm\rangle,i} (R)= 
  \begin{cases}
  2\text{Res}\{\mathcal{G}_{11}(R,E);E=E^{u}\} & \text{for }  + \\
    0 & \text{for }  - 
\end{cases}.                           
\end{equation}
This means that for the occupation to be nonzero the initial state must be symmetric with 
respect to the emitter's excitation, also implicitly revealing the symmetry of the APBSs 
above the band. 
Below the band, the situation is  different. We obtain
\begin{equation}
	\alpha^{l}_{|\pm\rangle,i} (R)= \text{Res} \left\{\mathcal{G}_{11}(R,E);E = E^{l} \right\} \left(1\pm \cos(\pi R) \right),
  \label{osci}
\end{equation}
showing that the occupation number oscillates on the length scale of the waveguide's 
site spacing. Indeed, the photonic part of the wave function changes its parity if 
one TLS is kept fixed and another moved from a given site to the nearest neighbour 
site. The difference in the behavior of the states below or above the band can be traced 
back to the different degeneracies of the band edges within the Brillouin zone: The upper 
band edge occurs at the single point $k=\pm\pi$ in the Brillouin zone, whereas the lower 
band edge occurs at $k=0$.
The F\"{o}rster potential corresponding to $|\Psi\rangle =|+\rangle$ is shown in Fig. \ref{FoersterCompPlotTot}.
Due to the parity cancellation mechanism, we observe a minimum at $R \approx 1.5$ which is 
much deeper than in the case of $|\Psi\rangle = |1\rangle$. Furthermore, there is no cusp in 
the $\phi_{|+\rangle}(R)$ since it coincides with a zero
in the occupation numbers given in Eq. \eqref{osci}. The oscillations we expect from Eq. \eqref{osci} 
fade out for $R > R^{l}_{\rm cr}$ due to destructive inference between the APBSs below the 
band edge. For the parameters we chose, the shifts $E^{u}_i(R)-E^{u}_i(\infty)$ above the 
band provide the dominating contribution to the potential (cf. Fig.~\ref{EigVal}). 

In summary, we have analyzed in detail the FRET interaction between two TLSs
mediated by the APBSs. The appearance of APBSs 
represents a non-Markovian effect which is a consequence of the boundedness of the 
waveguide spectrum. This behavior is absent in systems with unbounded 
dispersion relations, at least as long as the underlying rotating wave approximation is 
valid. Furthermore, we have demonstrated that the F\"{o}rster potential can change from 
repulsive to attractive (featuring minima and maxima), depending on the distance between 
the TLSs. The F\"{o}rster potential is also highly sensitive to the initial state of the 
system and we have compared two distinct initial states; in the first one, only one TLS
is excitated in the second one the excitation is distributed over the two TLSs in the 
form of an entangled state.
While the latter yield a pronounced minimum, the former features a cusp due to the sudden 
emergence of an additional APBS from the continuum of scattering states.
The structure of the F\"{o}rster potential of our model is much richer then one would expect 
from the standard three-dimensional case, where the potential is monotonous and of the 
typical dipole-dipole form $\sim R^{-3}$ \cite{Forster48,Milonni15}. This highlights the 
role of the 
electromagnetic environment which in our model is characterized by the bounded waveguide 
dispersion. The resulting novel features may lead to interesting applications. Firstly, 
the non-analytic behavior of $\phi_{|\Psi\rangle}(R)$ around $R_{\rm cr}$ facilitates the 
detection of a APBSs through suitably designed experiments. Secondly, the minima in the
F\"{o}rster potential can be used to trap atoms and/or nano-particles. Furthermore, it 
is quite conceivable to engineer a more interesting potential landscape by employing
more than two TLSs, as additional APBSs will become available. In particular,
this can be very interesting in the context of quantum simulations.

{\it Acknowledgments.} We acknowledge support by the Deutsche Forschungsgemeinschaft (DFG) through 
project B10 within the Collaborative Research Center (CRC) 951 Hybrid Inorganic/Organic Systems 
for Opto-Electronics (HIOS). FI acknowledge support from the DFG via the 
DIP program (grant SCHM 1049/7-1).



\newpage
\cleardoublepage
\setcounter{page}{1}

\newcounter{sfigure}
\setcounter{sfigure}{1}

\renewcommand{\theequation}{S\arabic{equation}}
\renewcommand{\thefigure}{S\arabic{sfigure}}

\setcounter{equation}{0}

\section*{{\large Supplemental Material}}
\subsection{\textsc{Nonperturbative and non-Markovian 
F\"{o}rster-interaction in waveguiding systems}}

\noindent 
In this supplemental material we provide further details on the calculations leading to 
the results presented in the main text.

\subsubsection{Green tensor formalism and occupation numbers}

The Green function of the system can be written as the ordered dyadic product of vectors 
containing the creation and the annihilation (bosonic and fermionic) operators involved in 
the system  (see the main text after Eq. (1)).
In the Heisenberg picture we have

\begin{widetext}
 \begin{equation}\label{eq:Appendix:GreensFunction1}
 \underline{\mathcal{G}}(t',t) = -i \left\langle 
  \begin{pmatrix}
   f^{\{1\}} (t) g^{\dagger \{1\}} (t) \\
   f^{\{2\}} (t) g^{\dagger \{2\}} (t) \\
   a_{k} (t)
  \end{pmatrix}
  \left( f^{\dagger \{1\}} (t') g^{\{1\}} (t'), f^{\dagger \{2\}}(t') g^{\{2\}}(t'), a_{k'}^{\dagger} (t') \right)
 \right\rangle,
\end{equation}
\end{widetext}
 or equivalently in the frequency domain as

\begin{equation}\label{eq:Appendix:GreensFunction2}
\underline{\mathcal{G}} (E) = 
 \begin{pmatrix}
  \mathcal{G}_{11} 		& \mathcal{G}_{21} 		& \mathcal{G}_{\text{ph},1} \\
  \mathcal{G}_{12} 		& \mathcal{G}_{22} 		& \mathcal{G}_{\text{ph},2} \\
  \mathcal{G}_{1,\text{ph}} 	& \mathcal{G}_{2,\text{ph}} 	& \mathcal{G}_{\text{w}} \\
 \end{pmatrix}.
\end{equation}
The Green functions with numbered indices appearing as elements of the above matrix 
describe an excitation moving from one TLS to another one (or staying at the same TLS), 
$\mathcal{G}_{\text{w}}$ describes photons in the waveguide - renormalized by interaction 
with both TLS - and the other Green functions describe absorption and emission processes.

The form of the Green tensor in Eq. (3) of the main text derives from the observation 
that, due to the Peierls substitution, the $R$-dependence in the perturbation series is 
totally encapsulated in the one-photon loop diagram (equivalent to the total system's 
self-energy) up to a possible phase factor due to external legs of the diagram. Its 
$3\times 3$ matrix structure is a consequence of the scattering processes (channels) 
characterizing our system in the one-photon sector. (See also the amputation rules for 
Feynman diagrams discussed in context of waveguide QED in Ref. \cite{Schneider16}.)
As an example of the calculation, we consider the Green's function $\mathcal{G}_{11}$. 
The corresponding Dyson equation reads 

\begin{equation}
\mathcal{G}_{11}(R)= 
 G_{1} +(U_1U_2)^2 \Gamma(R)  G_{1} \Gamma(R) G_{2} \mathcal{G} _{11}(R),
\end{equation}
where, for reasons of a compact notation, we have suppressed the argument $E$. This is
equivalent to
\begin{equation}
\mathcal{G}_{11}(R)= 
\frac{G_1}{1 -(U_1U_2)^2 \Gamma(R)^2 G_{2}G_{1}} \equiv \frac{\underline{[\mathcal{M}}(R)]_{11}}{Q(R)},
\label{TLSGreensfunction}
\end{equation}
where $G_{1}$ and $G_{2}$ are the, $R-$independent, one-scatterer renormalized Green 
functions \cite{Schneider16}. 
We observe that the $R-$dependent self-energy, $\Gamma(R)$ as given in the main text, 
does not depend on $k$, since the photon momentum is integrated out. This explains the 
factorization mentioned in the main text.

The same Green tensor formalism can be used to extract the occupation numbers  
$\alpha_{|\Psi\rangle,i}=|\langle \Psi| E_i \rangle|^2$. Together with the eigenenergies, 
they are one of the two main elements involved in the definition of the F\"{o}rster 
potential in Eq. (2) of the main text.

We start by noticing that to each Green tensor we can associate a quantum operator.
For example the waveguide ($w$) Green function denoted by $\mathcal{G}_{w}$ is given in Heisenberg 
representation by $\mathcal{G}_{w}=\langle 0|a_k a_k^{\dag} |0\rangle $ 
(cf. Eqs.~\eqref{eq:Appendix:GreensFunction1} and \eqref{eq:Appendix:GreensFunction2}) 
and we can identify $ \hat{G}_{w}=\hat{a}_k \hat{a}_k^{\dag}$ as the corresponding 
quantum operator. In general, it is convenient to express this operator using a spectral 
decomposition (often called Lehmann representation in many-body theory  \cite{Bruus04})

\begin{equation}
 \hat{G}(E)=\sum_i \frac{|E_i \rangle \langle E_i|}{E-E_i},
\end{equation} 
where $\{|E_i \rangle\}$ corresponds to the energy eigenbasis.
For a general state $|\Psi\rangle$ we then have 
\begin{equation}
\langle \Psi| \hat{G}(E)|\Psi\rangle=\sum_i \frac{|\langle \Psi|E_i\rangle|^2}{E-E_i}.
\end{equation}
Invoking the residue theorem and integrating around a closed contour which contains exactly one eigenvalue, we have
\begin{equation}\label{OccviaGreen}
\mathrm{Res}(\langle \Psi|\hat{G}(E)|\Psi\rangle,E=E_i)=|\langle \Psi|E_i\rangle|^2 \equiv\alpha_{|\Psi\rangle,i}~.
\end{equation}

Let us apply the above general considerations to one of the configurations analyzed in 
the main text. Consider a system of a waveguide coupled to two TLSs, one of which is 
excited, i.e. $|\Psi\rangle=|1\rangle = |\uparrow,\downarrow,0 \rangle$. In this case 
the relevant Green tensor is $\langle 1|\hat{G}(E)|1\rangle=\mathcal{G}_{11}(R,E)$, 
which is the Green function already defined in Eq.\eqref{TLSGreensfunction}. This means
\begin{align}
\label{Resoneup}
\alpha_{|1\rangle,i}= \mathrm{Res}\{\mathcal{G}_{11}(E),E=E_i\}, 
\end{align}
where $E_i$ are the solutions for $|E_i|>2$ of Eq. (4) of the main text.
\subsubsection{Asymptotic expressions for the APBSs energies}

In following section we want to motivate the asymptotic values for the eigenenergies 
of the system given in Fig. 6 of the main text.
We employ the parametrization $\Theta=\mathrm{arccosh}(E/2)$ as described in the main 
text and for simplicity let us define 
\begin{equation}
f(\Theta)=\sinh(2\Theta)-\Omega \sinh(\Theta)-\frac{U^2}{2},
\end{equation}
focusing on the regime $E^{u}>2$ removing the corresponding superscript. The values below 
the band ($E^{l}_{\pm}$) can be obtained by making the replacement $R\to -R$ and 
$\Omega \to-\Omega$ in all following equations. 
For $R\Theta_{0,\pm} \ll1 $, where $\Theta_{0,\pm}$ are the values corresponding to 
$E_{\pm}(R\to 0)$, we write $\Theta=\Theta_{0,\pm}+\delta_{0,\pm}(R)$ with $\delta/\Theta_0\ll1$ 
and get from Eq. (5) in the main text:

\begin{equation}
\label{ThetaSmallREq}
\delta_{0,\pm}(R) \approx 
\mp\frac{U^2}{2}\frac{R\Theta_{0,\pm} }{2 \cosh (2 \Theta_{0,\pm})-2 \Omega \cosh (\Theta_{0,\pm})},
\end{equation}
where we have neglected terms of order $R\delta(R)$, since they turn out to be of order $\mathcal{O}(R^2)$.  
Expanding $E_{\pm}=2\cosh(\Theta_{0,\pm}+\delta_{0,\pm}(R))$ up to first order we have
\begin{equation}\label{ERsmall}
E_{\pm}\sim2 \cosh(\Theta_{0,\pm})\mp\gamma_{0,\pm}R,
\end{equation}
which are the expressions that are reported in the plots of Fig. 3 of the main text.
For brevity, we have defined the term $\gamma_{0,\pm}=\sinh(\Theta_{0,\pm}) U^2 /f'(\Theta_{0,\pm})$ 
(the prime denotes a derivative with respect to the argument).

We use a similar approach for the opposite limit $\Theta_{\infty,\pm} R\gg1$, where $\Theta_{\infty,\pm}$ corresponds to the eigenenergy $E_{\pm}(R\to \infty)$. Solving self-consistently Eq. (5) of the main text, we obtain
\begin{equation}
\delta_{\infty,\pm}(R) \sim \pm \frac{  U^2e^{-\Theta_{\infty}R}}{f'(\Theta_{\infty})}
\end{equation}
(in this case $ \exp(-\Theta R)/2 \ll 1$ is the small quantity relevant for the expansion).
This yelds
\begin{equation}\label{EnasyRinfmt}
E_{\pm}\sim2 \cosh(\Theta_{\infty})\pm \underbrace{\frac{\sinh(\Theta_{\infty})U^2}{f'(\Theta_{\infty})}}_{= \gamma_{\infty}}e^{-R\Theta_{\infty}},
\end{equation}
which, again, are the expressions that are reported in Fig. 3 of the main text. 
Equation \eqref{EnasyRinfmt} shows how the energies become degenerate for large 
separations while they progressively split (level repulsion) as the TLSs get 
closer.

Finally, for $R\approx R_{\rm cr}$ we might perform an expansion in $\Delta R =R-R_{\rm cr}$, 
obtaining ($\Delta R /R_{\rm cr}\ll1$) 
\begin{equation}
\Delta R-\Theta\frac{R^2}{2}\approx 0,
\end{equation}
which and after some simple algebra leads to the result stated in the main text, i.e.  
\begin{align}\label{EnasyRcmt}
E\sim 2+\underbrace{\frac{4}{R_c^2} }_{\gamma_{cr}}(\Delta R)^2.
\end{align}
Notice that the case where $R_{\rm cr} \to 0$ features additional complications that 
require a higher-order expansion resulting in
\begin{equation}
E\approx 2+\frac{U \sqrt{3\Delta R} }{\sqrt{8-\Omega}}.
\end{equation}

\vspace{0.5cm}

\noindent T. Sproll$^{1}$, Ch. Martens$^{1}$, M. P. Schneider$^{1}$, F. Intravaia$^{1,2}$ and K. Busch$^{1,2}$.

\begin{small}
\begin{enumerate}
\item[$^{1}$]
Max-Born-Institut, 12489 Berlin, Germany
\item[$^{2}$]
Humboldt-Universit\"at zu Berlin, Institut f\"ur Physik, AG Theoretische Optik \& Photonik, 12489 Berlin, Germany
\end{enumerate}
\end{small}


\begin{thebibliography}{10}

\bibitem{Forster48}
T. F\"{o}rster, Annalen der Physik {\bf 437},  55  (1948).

\bibitem{Sekar03}
R.~B. Sekar and A. Periasamy, J. Cell Biol. {\bf 160},  629  (2003).

\bibitem{Saini06}
S. Saini, H. Singh, and B. Bagchi, J. Chem. Sci. {\bf 118},  23  (2006).

\bibitem{Endo10}
I. Endo and T. Nagamune, {\em Nano/Micro Biotechnology} (Springer, 
  2010).

\bibitem{Pollock99}
B.~A. Pollock and R. Heim, Trends in Cell Biology {\bf 9},  57  (1999).

\bibitem{Truong01}
K. Truong and M. Ikura, Current Opinion in Structural Biology {\bf 11},  573
  (2001).

\bibitem{Ishikawa12}
H.~C. Ishikawa-Ankerhold, R. Ankerhold, and G.~P. Drummen, Molecules {\bf 17},
  4047  (2012).

\bibitem{Longo11}
P. Longo, P. Schmitteckert, and K. Busch, Phys. Rev. A {\bf 83},  063828
  (2011).

\bibitem{Zheng13}
H. Zheng and H.~U. Baranger, Phys. Rev. Lett. {\bf 110},  113601  (2013).

\bibitem{Laakso14}
M. Laakso and M. Pletyukhov, Phys. Rev. Lett. {\bf 113},  183601  (2014).

\bibitem{Shahmoon14}
E. Shahmoon, I. Mazets, and G. Kurizki, P.N.A.S.  (2014).

\bibitem{Hood16}
J.~D. Hood {\it et~al.}, P.N.A.S. {\bf 113},  10507  (2016).

\bibitem{Shi16}
T. Shi, Y.-H. Wu, A. Gonz\'alez-Tudela, and J.~I. Cirac, Phys. Rev. X {\bf 6},
  021027  (2016).

\bibitem{Lodahl17}
P. Lodahl {\it et~al.}, Nature {\bf 541},  473  (2017).

\bibitem{Ladd10}
T.~D. Ladd, F. Jelezko, R. Laflamme, Y. Nakamura, C. Monroe, and J.~L. O'Brien,
  Nature {\bf 464},  45  (2010).

\bibitem{Kimble08}
H.~J. Kimble, Nature {\bf 453},  1023  (2008).

\bibitem{Reitz13}
D. Reitz, C. Sayrin, R. Mitsch, P. Schneeweiss, and A. Rauschenbeutel, Phys.
  Rev. Lett. {\bf 110},  243603  (2013).

\bibitem{Arcari14}
M. Arcari {\it et~al.}, Phys. Rev. Lett. {\bf 113},  093603  (2014).

\bibitem{John90}
S. John and J. Wang, Phys. Rev. Lett. {\bf 64},  2418  (1990).

\bibitem{Auerbach94}
A. Auerbach, {\em Interacting Electrons and Quantum Magnetism} (Springer, 1994).

\bibitem{Schneider16}
M.~P. Schneider, T. Sproll, C. Stawiarski, P. Schmitteckert, and K. Busch,
  Phys. Rev. A {\bf 93},  013828  (2016).

\bibitem{Mercouris96}
T. Mercouris, Y. Komninos, S. Dionissopoulou, and C.~A. Nicolaides, J. Phys. B:
  At. Mol. Opt. Phys. {\bf 40},  2133  (1997).

\bibitem{Feranchuk96}
I. Feranchuk, I.~I. Komarov, and A.~P. Ulyanenkov, J. Phys. A: Math. Gen. {\bf
  29},  4035  (1996).

\bibitem{Stillinger75}
F.~H. Stillinger and D.~R. Herrick, Phys. Rev. A {\bf 93},  446  (1975).

\bibitem{Zumofen08}
G. Zumofen, N.~M. Mojarad, V. Sandoghdar, and M. Agio, Phys. Rev. Lett. {\bf
  101},  180404  (2008).

\bibitem{Sproll16}
T. Sproll, Ph.D. thesis, Humboldt-Universit\"{a}t zu Berlin, 2016.

\bibitem{Note1}
For a general dispersion relation $\epsilon (k)$ and two identical emitters, one can show some real solutions of Eq.  (\ref {GeneralResCond}),
which provides the eigenenergies of the Hamiltonian in Eq. (\ref{eq:Hamiltonian}),
always occur for 
  \begin {equation*}  
  E_{\protect \rm BIC} = \Omega = \epsilon \left ( \protect \frac {\pi N}{R} \right ), \quad N\in \mathbb {Z}. 
    \label {GeneralBIC}
  \end {equation*} 
The solutions of the previous equation do not depend on the atom-waveguide coupling physics parameters and are located
within the continuum of the band energy spectrum (BIC). Although in Fig. \ref{EnergySpectrum} several BICs are represented, depending on the emitters' separation, only one state is selected by the system. Specifically, for the BIC obtained in the limit $R \to 0$ the system's dynamics shows some similarities with the behavior of the Dark States considered in C. Martens, P. Longo, and K. Busch, New J. Phys. {\bf 15}, 083019  (2013). Indeed, for $R \to 0$ in the one excitation sub-space our configuration turns out to be equivalent to a V-system.
For non-identical  TLSs, the cavity formed by the two emitters becomes leaky due to a mismatch in the individual mirror characteristics.

\bibitem{Economou05}
E.~N. Economou,  in {\em Green's functions in quantum physics}, {\em
  Solid-state sciences}, edited by M. Cardona, P. Fulde, K. von Klitzing, R.
  Merlin, and H.-J. Q.~H. St{\"o}rmer (Springer, Berlin / Heidelberg / New
  York, 2005).

\bibitem{Liu17}
Y. Liu and A.~A. Houck, Nat. Phys. {\bf 13},  48  (2017).

\bibitem{Casimir48}
H.~B.~G. Casimir, Proc. K. Ned. Akad. Wet. {\bf 51},  793  (1948).

\bibitem{Intravaia05}
F. Intravaia and A. Lambrecht, Phys. Rev. Lett. {\bf 94},  110404  (2005).

\bibitem{Intravaia07}
F. Intravaia, C. Henkel, and A. Lambrecht, Phys. Rev. A {\bf 76},  033820
  (2007).

\bibitem{Intravaia12b}
F. Intravaia and R. Behunin, Phys. Rev. A {\bf 86},  062517  (2012).

\bibitem{Wylie85}
J.~M. Wylie and J.~E. Sipe, Phys. Rev. A {\bf 32},  2030  (1985).

\bibitem{Intravaia11}
F. Intravaia, C. Henkel, and M. Antezza,  in {\em Casimir Physics}, Vol.~834 of
  {\em Lecture Notes in Physics}, edited by D. Dalvit, P. Milonni, D. Roberts,
  and F. da~Rosa (Springer, Berlin / Heidelberg, 2011), pp.\ 345--391.

\bibitem{Laliotis14}
A. Laliotis, T.~P. de~Silans, I. Maurin, M. Ducloy, and D. Bloch, Nat. Commun.
  {\bf 5},    (2014).

\bibitem{Haakh12b}
H.~R. Haakh, J. Schiefele, and C. Henkel, Int. J. Mod.Phys.: Conf. Ser. {\bf
  14},  347  (2012).

\bibitem{Donaire15a}
M. Donaire, R. Gu{\'e}rout, and A. Lambrecht, Phys. Rev. Lett. {\bf 115},
  033201  (2015).

\bibitem{Milonni15}
P.~W. Milonni and S.~M.~H. Rafsanjani, Phys. Rev. A {\bf 92},  062711  (2015).

\bibitem{Kocabas16}
S.~E. Kocaba\c{s}, Phys. Rev. A {\bf 93},  033829  (2016).

\bibitem{Mahan00}
G.~D. Mahan, {\em Many-Particle Physics} (Springer,  2000).

\bibitem{Weisskopf30}
V. Weisskopf and E. Wigner, Z. Physik {\bf 63},  54  (1930).

\bibitem{Berman10}
P.~R. Berman and G.~W. Ford,  in {\em Advances In Atomic, Molecular, and
  Optical Physics}, edited by E. Arimondo, P.~R. Berman, and C.~C. Lin
  (Academic Press, Amsterdam, 2010), Vol.~59, p.\ 175.

\bibitem{SuppMat}
Supplemental Material at [URL will be inserted by pub- lisher].


\bibitem{Haakh13}
H.~R. Haakh and F. Intravaia, Phys. Rev. A {\bf 88},  052503  (2013).

\end{thebibliography}

\begin{thebibliography}{1}

\bibitem{Schneider16}
M.~P. Schneider, T. Sproll, C. Stawiarski, P. Schmitteckert, and K. Busch, {\em
  "{G}reen's-function formalism for waveguide {QED} applications"}, Phys. Rev.
  A {\bf 93},  013828  (2016).

\bibitem{Bruus04}
H. Bruus and K. Flensburg, {\em Many-body quantum theory in condensed matter
  physics} (Oxford University Press Oxford, New York, 2004).

\end{thebibliography}

\end{document}